\documentclass{article}
\usepackage{amsmath}    
\usepackage{graphicx,dcolumn,amsmath,amssymb,amsfonts,epsfig,float,subfigure}
\usepackage{amsthm}
\newtheorem{theorem}{Theorem}

\numberwithin{equation}{section}

\begin{document}

\begin{center}

{\Large Third order superintegrable systems separating in polar coordinates }

\vskip 0.7cm

Fr\'ed\'erick Tremblay\footnote{tremblaf@crm.umontreal.ca} and Pavel Winternitz\footnote{wintern@crm.umontreal.ca},\\
Centre de recherches math\'ematiques and D\'epartement de math\'ematiques et de
statistique,
Universit\'e de Montreal,  C.P. 6128,
succ. Centre-ville, Montr\'eal (QC) H3C 3J7, Canada \\[10pt]

\end{center}

\vskip .9cm

\begin{abstract}
A complete classification is presented of quantum and classical superintegrable systems in $E_2$ that allow the separation of variables in polar coordinates and admit an additional integral of motion of order three in the momentum. New quantum superintegrable systems are discovered for which the potential is expressed in terms of the sixth  Painlev\'e transcendent  or in terms of the Weierstrass elliptic function.
\end{abstract}

\section{Introduction}
The purpose of this article is to obtain and classify all classical and quantum Hamiltonians $H$  that allow the separation of variables in polar coordinates and admit a third order integral of motion $Y$. The system under study is characterized by three conserved quantities:
 \begin{align}
 \label{H}
 H&=\frac{p_1^2+p_2^2}{2}+V(r,\theta)\\
\label{X}
 X&=L_3^2+2S(\theta)\\
\label{Y}
 Y&=\sum_{i+j+k=3}A_{ijk}\{L_3^i,p_1^jp_2^k\}+\{g_1(x,y),p_1\}+\{g_2(x,y),p_2\}
 \end{align}
where 
\begin{equation}
\label{pot}
V(r,\theta)=R(r)+\frac{S(\theta)}{r^2}
\end{equation}
Here $R(r)$ and $S(\theta)$ are arbitrary functions and $A_{ijk}$ are real constants. The polar coordinates are defined as usual: $x=r\cos\theta$ and $y=r\sin\theta$.

In classical mechanics $p_1,p_2$ are the cartesian components of linear momentum and $L_3$ is the two-dimensional angular momentum. In quantum mechanics we have
\begin{equation}
p_1=-i\hbar\frac{\partial}{\partial x},\quad p_2=-i\hbar\frac{\partial}{\partial y},\quad L_3=-i\hbar\frac{\partial}{\partial \theta}
\end{equation}
The curly brackets, $\{\cdot,\cdot\}$, in (\ref{Y}) denote an anticommutator in quantum mechanics. In classical mechanics we have $\{L_3^i,p_1^jp_2^k\}=2L_3^ip_1^jp_2^k$.

This study is part of a general program devoted to superintegrable systems in classical and quantum mechanics. Roughly speaking, an integrable system is superintegrable if it allows more integrals of motion than degrees of freedom. For more precise definitions and an extensive bibliography see e.g. \cite{mar1,win1}. A system in $n$ dimensions is maximally superintegrable if it allows $2n-1$ integrals. In classical mechanics the integrals of motion must be well defined and functionally independent functions on phase space and typically at least one subset of $n$ integrals (including the Hamiltonian) is in involution. In quantum mechanics the integrals should be well-defined linear operators in the enveloping algebra of the Heisenberg algebra with basis $\{x_i, p_i,\hbar\}$ for $i=1, ..., n$ and they should be algebraically independent (within the Jordan algebra generated by their anticommutators).

The majority of publications on superintegrability is devoted to the quadratic case when the integral of motion is quadratic  in the momenta (see e.g. \cite{winfris1}-\cite{mil2}). Quadratic superintegrability for one particle in a scalar potential is related to multiseparability in the Schr\"odinger equation or the Hamilton-Jacobi equation, in quantum or classical mechanics, respectively.

More recently some of the interest has shifted to higher order integrability. An infinite family of superintegrable and exactly solvable systems in a Euclidean plane has been proposed \cite{ttw2,ttw3}. The potential depends on a real number $k>0$. It has been conjectured that this sytem is superintegrable in quantum mechanics for all integer values of $k$ with one integral of order 2 and the other of order $2k$. For brevity we will call it the TTW model. The superintegrability of the TTW model  has been so far confirmed for odd values of $k\geq3$ \cite{quesne}. In the classical case all bounded trajectories are periodic \cite{ttw3} for all integer and rational values of $k$ and superintegrability has been proven for such values of $k$ \cite{mil3}. Both the trajectories and the higher order integral of motion can be expressed in terms of Chebyshev polynomials. The generalisation of the TTW model  in a three dimensional Euclidean space has been recently proposed \cite{mil4}.

A systematic study of integrable systems in classical mechanics with one third order integral of motion was initiated by Drach in a remarkable paper published in 1935 \cite{Drach}. He considered classical Hamiltonian mechanics in a two dimensional complex Euclidean plane and found 10 potentials allowing a third order integral. More recently it was shown that 7 of these systems are actually quadratically superintegrable and that the third order integral of motion is a Poisson commutator of two independent second order ones \cite{ran,tsi}. Quadratically integrable (and superintegrable) potentials coincide in classical and quantum mechanics. This is not necessarily the case when higher order integrals are involved \cite{Hiet1,Hiet2}.

A systematic search for quantum and classical superintegrable systems in a real Euclidean plane  with one third order integral of motion and one first order or second order one was started in \cite{gravel1,gravel2}. In \cite{gravel2} the second order integral of motion was chosen so as to assure separation of variables in cartesian coordinates. This lead to several new classical superintegrable systems but mainly to completely new quantum ones, in which the potential is expressed in terms of Painlev\'e transcendents. The integrals of motion generate polynomial algebras \cite{dask}, \cite{grav}-\cite{mar7}. Their representation theory was used to calculate the energy spectra and a relation with supersymmetric quantum mechanics was used to calculate wave functions \cite{mar3}-\cite{mar7}.

The existence of a third order integral of motion in quantum mechanics was investigated earlier \cite{data} and potentials expressed in terms of the Weierstrass function were obtained.

In this article we continue with the classification of superintegrable systems and impose the conditions (1.1) to (1.4), i.e. separation of variables in polar coordinates.

The determining equations for the existence of a third order integral are presented in Section 2 in polar coordinates. The possible form of the radial part of the potential is established in Section 3 and summed up in Theorem 1. The angular part is discussed in Section 4. Genuinely new superintegrable potentials are obtained when the radial part of the potential vanishes. The angular parts of the potential are expressed in terms of the sixth Painlev\'e transcendent $P_6$ or in terms of the Weierstrass elliptic function. The main results are summed up as theorems in Section 5.
\section{Determining equations of a third order integral of motion in polar coordinates}

The quantities $X$ and $H$ commute in quantum mechanics and Poisson-commute in the classical case. The form of (\ref{Y}) assures that all terms of order 4 and 3 in the (Poisson) commutator $[H,Y]=0$
vanish. The vanishing of lower order terms provides the following determining equations for the functions $g_1,g_2$ and $V$ in (\ref{H}) and (\ref{Y}):
\begin{align}
\label{detunpol}
 G_1V_r+G_2V_{\theta} &= \frac{\hbar^2}{4}\Bigg[F_1V_{rrr}+F_2V_{rr\theta}+F_3V_{r\theta\theta}+F_4V_{\theta\theta\theta}+rF_3V_{rr}+ \Big(3rF_4 \nonumber\\&- \frac{2}{r}F_2\Big)V_{r\theta} 
 -\frac{2}{r}F_3V_{\theta\theta}+ (-F_3 + 2C_{1}\cos\theta + 2C_{2}\sin\theta)V_r\nonumber\\
&+ \Big(-2F_4 +\frac{2}{r^2}F_2 +8D_{0} +
\frac{(-2C_{1}\sin\theta +
2C_{2}\cos\theta)}{r}\Big)V_{\theta}\Bigg]
\end{align}

\begin{align}
\label{detdeuxpol}
(G_1)_r &= 3F_1V_r+F_2V_{\theta}\\
\label{dettroispol}
\frac{(G_2)_{\theta}}{r^2}&= F_3V_r + 3F_4V_{\theta} - \frac{G_1}{r^3}\\
\label{detquatrepol} 
(G_2)_r&= 2(F_2V_r
+F_3V_{\theta})-\frac{(G_1)_{\theta}}{r^2}
\end{align}
with:
\begin{align}
\label{fonctionun}
F_1(\theta) &= A_{1}\cos3\theta + A_{2}\sin3\theta +A_{3}\cos\theta + A_{4}\sin\theta,\\
\label{fonctiondeux}
F_2(r,\theta)&=\frac{1}{r}(-3A_{1}\sin3\theta + 3A_{2}\cos3\theta -A_{3}\sin\theta + A_{4}\cos\theta)\nonumber\\
&+ B_1\cos2\theta + B_2\sin2\theta + B_0,
\\
\label{fonctiontrois}
F_3(r,\theta)&=\frac{1}{r^2}(-3A_{1}\cos3\theta - 3A_{2}\sin3\theta +A_{3}\cos\theta + A_{4}\sin\theta)\nonumber\\
& +\frac{1}{r}(-2B_1\sin2\theta + 2B_2\cos2\theta)\nonumber\\
 &+C_1\cos\theta + C_2\sin\theta,
\\
\label{fonctionquatre}
F_4(r,\theta)&=\frac{1}{r^3}(A_{1}\sin3\theta - A_{2}\cos3\theta -A_{3}\sin\theta + A_{4}\cos\theta)\nonumber\\
& +\frac{1}{r^2}(-B_1\cos2\theta - B_2\sin2\theta + B_0)\nonumber\\
 &+\frac{1}{r}(-C_1\sin\theta + C_2\cos\theta) + D_0,
\end{align}
and 
\begin{gather*}
G_1(r,\theta) = g_1\cos\theta  + g_2\sin\theta, \qquad
G_2(r,\theta) = \frac{-g_1\sin\theta  + g_2\cos\theta}{r}
\end{gather*}

The constants $A_i$, $B_i$, $C_i$ et $D_0$ are related to 
$A_{ijk}$ of (\ref{Y}):
\begin{gather*}
 A_1=\frac{A_{030}-A_{012}}{4},\quad A_2
=\frac{A_{021}-A_{003}}{4}, \quad A_3
=\frac{3A_{030}+A_{012}}{4} \\
 A_4
=\frac{3A_{003}+A_{021}}{4},\quad B_1 = \frac{A_{120}-A_{102}}{2},\quad B_2 = \frac{A_{111}}{2},
 \\B_0= \frac{A_{120}+A_{102}}{2},\quad C_1 = A_{210},\quad C_2=A_{201},\quad D_0=A_{300}
\end{gather*}
The constants  $A_i$, $B_i$, $C_i$ et $D_0$ are real. Relations equivalent to (\ref{detunpol})-(\ref{detquatrepol}) were already obtained in \cite{gravel1} in cartesian coordinates.

The third order constant of motion can be written in terms of  $A_i$, $B_i$, $C_i$ and $D_0$ as
\begin{align}
\label{intmod}
 Y&=D_0L_3^3+C_1\{L_3^2,p_1\}+C_2\{L_3^2,p_2\}+B_0L_3(p_1^2+p_2^2)+B_1\{L_3,p_1^2-p_2^2\}\nonumber\\&+2B_2\{L_3,p_1p_2\}+A_1p_1(p_1^2-3p_2^2)+A_2p_2(3p_1^2-p_2^2)+(A_3p_1+A_4p_2)(p_1^2+p_2^2)\nonumber \\&+\{g_1(x,y),p_1\} + \{g_2(x,y),p_2\}
\end{align}

Under rotations the Hamiltonian (\ref{H}) and the integral $X$ (\ref{X}) remain invariant but the third order integral $Y$ (\ref{Y}) or (\ref{intmod}) transforms into a new integral of the same form with new coefficients and new functions $g_1,g_2$. The constants $D_0$ and $B_0$ are singlets under rotations and hence invariants. The expressions $\{B_1,B_2\}$,  $\{A_1,A_2\}$ and $\{A_3,A_4\}$ are doublets under rotations.

The existence of the third order integral $Y$ depends on the compatibility of (\ref{detunpol})-(\ref{detquatrepol}). Equation (\ref{detunpol}) establishes the difference between the quantum and the classical cases. We obtain the classical analog of (\ref{detunpol}) by setting $\hbar\mapsto0$.

From (\ref{detdeuxpol})-(\ref{detquatrepol}) we can deduce a compatibility condition for the potential $V(r,\theta)$, namely the third order linear differential equation:
\begin{eqnarray}
\label{compa}
0=r^4F_3V_{rrr}+\Big(3r^4F_4-2r^2F_2\Big)V_{rr\theta}+\Big(3F_1-2r^2F_3\Big)V_{r\theta\theta}+F_2V_{\theta\theta\theta} \nonumber\\
+\Big(2r^4F_{3r}+6r^3F_3-2r^2F_{2\theta}-3rF_1\Big)V_{rr}+\Big(2F_{2\theta}-4rF_3-2r^2F_{3r}\Big)V_{\theta\theta} \nonumber\\
+\Big(6r^4F_{4r}+18r^3F_4-2r^2(F_{2r}+F_{3\theta})-5rF_2+6F_{1\theta}\Big)V_{r\theta} \nonumber\\
+\Big(r^4F_{3rr}+6r^3F_{3r}+r^2(6F_3-2F_{2r\theta})-4rF_{2\theta}+3F_{1\theta\theta}\Big)V_r \nonumber\\
+\Big(3r^4F_{4rr}+18r^3F_{4r}+r^2(18F_4-2F_{3r\theta})-r(F_{2r}+4F_{3\theta})+F_{2\theta\theta}\Big)V_\theta
\end{eqnarray}
Compatibility of (\ref{detdeuxpol})-(\ref{detquatrepol}) and (\ref{detunpol}) imposes further conditions on $V(r,\theta)$, they are however nonlinear \cite{gravel1}.

Our aim is to find all solutions of (\ref{detunpol})-(\ref{detquatrepol}) and thereby find all superintegrable classical and quantum systems that separate in polar coordinates and allow a third order integral $Y$. The existence of $X$ (\ref{X}) is guaranteed by the form (\ref{pot}) of the potential and is directly related to the separation of variables in polar coordinates.

Specifying the radial dependance of (\ref{fonctiondeux}) to (\ref{fonctionquatre}):
\begin{align}
\label{eq:F2}
F_2(r,\theta) &= \frac{1}{r}F_{21}(\theta) + F_{20}(\theta)\\
\label{eq:F3}
F_3(r,\theta) &= \frac{1}{r^2}F_{32}(\theta)+\frac{1}{r}F_{31}(\theta)+F_{30}(\theta)\\
\label{eq:F4} F_4(r,\theta) &=
\frac{1}{r^3}F_{43}(\theta)+\frac{1}{r^2}F_{42}(\theta)+\frac{1}{r}F_{41}(\theta)
+ D_0
\end{align}
we can deduce:
\begin{equation}
\label{G1} G_1(r,\theta) = 3F_1\Big(R + \frac{1}{r^2}S\Big) -
\Big(\frac{F_{21}}{2r^2} + \frac{F_{20}}{r}\Big)\dot{S} +
\beta(\theta)
\end{equation}
where $\dot{S}(\theta)=\displaystyle\frac{d}{d\theta}S(\theta)$. The derivative with respect to $r$ will be denoted by a prime so that $R'(r)=\displaystyle\frac{d}{dr}R(r)$.
\section{Radial term in the potential}

With the separation of the potential in polar coordinates (\ref{pot}), (\ref{compa}) can be expressed
as:
\begin{equation}
\label{comptrois1sep} m_3R^{(3)}+m_2R''+m_1R' +
n_3S^{(3)}+n_2\ddot{S}+n_1\dot{S}+n_0S=0
\end{equation}
for $m_i=m_i(r,\theta)$ and $n_i=n_i(r,\theta)$ in terms of the functions $F_i(r,\theta)$ and their derivatives. The $n_i=n_i(r,\theta)$ are linear in $r$ : $n_i = n_{i0}(\theta)
+n_{i1}(\theta)r$. In this way, by differentiating  (\ref{comptrois1sep}) two times with respect to $r$, we obtain a linear ordinary differential equation for $R(r)$:
\begin{align}
\label{dercompun}  m_3R^{(5)}+(2m_{3r} +
m_2)R^{(4)}+(m_{3rr}&+2m_{2r}+m_1)R^{(3)}\nonumber\\&+(m_{2rr}+2m_{1r})R''+m_{1rr}R'=0
\end{align}
In this expression, by using the form of the  $F_i(r,\theta)$, the coefficients associated to the derivatives of different orders of $R(r)$ can be expressed as combinations of trigonometric functions and different powers of $r$. Thus,  by separating  (\ref{dercompun}), according to the different linearly independent trigonometric functions,  we obtain six differential equations  for $R(r)$ that have to vanish independently of each other:
\begin{align}
\label{r1}
&A_1(r^4R^{(5)}+7r^3R^{(4)}-r^2R^{(3)}-18rR''+18R')=0\\
\label{r2}
&A_2(r^4R^{(5)}+7r^3R^{(4)}-r^2R^{(3)}-18rR''+18R')=0\\
\label{r3}
&B_1(r^4R^{(5)}+14r^3R^{(4)}+48r^2R^{(3)}+24rR''-24R')=0\\
\label{r4}
&B_2(r^4R^{(5)}+14r^3R^{(4)}+48r^2R^{(3)}+24rR''-24R')=0\\
\label{r5}
& \Big(C_1r^6+A_3r^4\Big)R^{(5)}+\Big(20C_1r^5+11A_3r^3\Big)R^{(4)}
+\Big(120C_1r^4+27A_3r^2\Big)R^{(3)}\nonumber\\&+\Big(240C_1r^3+6A_3r\Big)R''
+\Big(120C_1r^2-6A_3r\Big)R'=0\\
\label{r6}
&\Big(C_2r^6+A_4r^4\Big)R^{(5)}+\Big(20C_2r^5+11A_4r^3\Big)R^{(4)} +\Big(120C_2r^4+27A_4r^2\Big)R^{(3)}\nonumber\\&+\Big(240C_2r^3+6A_4r\Big)R''
+\Big(120C_2r^2-6A_4r\Big)R'=0
\end{align}

The solutions of these equations are:
\begin{enumerate}
 \item $A_1^2+A_2^2\neq0$\\
\begin{equation}
 \label{RAi}
 R(r)= \frac{a_3}{r}+a_4r^2+a_5r^4
\end{equation}
\item $B_1^2+B_2^2\neq0$\\
\begin{equation}
  \label{RBi}
 R(r)=\frac{a_2}{r^3}+\frac{a_3}{r}+a_4r^2
\end{equation}
\item $C_1^2+C_2^2\neq0$\\
\begin{equation}
  \label{RCi}
 R(r)=\frac{a_1}{r^4}+\frac{a_2}{r^3}+\frac{a_3}{r}
\end{equation}
\item $A_3^2+A_4^2\neq0$\\
\begin{equation}
 \label{RAAi}
 R(r)=\frac{a_3}{r}+a_4r^2+a_6\log r
\end{equation}
\item $(C_1A_3)\neq0$\\
\begin{equation}
 \label{RAA3C1}
 R(r)=\frac{a_3}{r}+\frac{a_7}{\sqrt{A_3+C_1r^2}}+\frac{a_8}{\sqrt{A_3+C_1r^2}}\log\Bigg(\frac{\sqrt{A_3}+\sqrt{A_3+C_1r^2}}{r}\Bigg)
\end{equation}
\item $(C_2A_4)\neq0$\\
\begin{equation}
  \label{RAA4C2}
 R(r)=\frac{a_3}{r}+\frac{a_7}{\sqrt{A_4+C_2r^2}}+\frac{a_8}{\sqrt{A_4+C_2r^2}}\log\Bigg(\frac{\sqrt{A_4}+\sqrt{A_4+C_2r^2}}{r}\Bigg)
\end{equation}
\end{enumerate}
We omit $\frac{1}{r^2}$  and the constant terms since they can be absorbed into the angular part.

\begin{theorem}
A third order integral of motion for a potential of the form (\ref{pot}) must have the form (\ref{Y}) and can exist only if one of the following situations occurs:
\begin{enumerate}
\item The constants in (\ref{intmod}) satisfy
\begin{equation}A_1=A_2=A_3=A_4=B_1=B_2=C_1=C_2=0\end{equation}
Then the radial equations (\ref{r1}) to (\ref{r6})  impose no restriction on $R(r)$.
\item $R(r)=\frac{a}{r}$ for $a\neq0$. 
\item $R(r)=ar^2$ for $a\neq0$ and $C_1=C_2=0$.
\item $R(r)=0$.
\end{enumerate}

In cases (ii) and (iv) equations (\ref{r1}) to (\ref{r6}) are satisfied identically for all values of the constants in (\ref{Y}) or (\ref{intmod}).
\end{theorem}

Before prooving Theorem 1 let us stress that it gives only necessary conditions for the existence of the integral $Y$, not sufficient ones. Those will be obtained when integrating the equations for the angular part $\frac{S(\theta)}{r^2}$.
\begin{proof}
(i) Let us start with Case (\emph{i}) of the theorem. The condition (3.15) implies that (\ref{r1}) to (\ref{r6}) are satisfied identically for all $R(r)$. Then (2.5) to (2.8) simplify to
\begin{equation*}
F_1=F_3=0,\quad F_2=B_0,\quad F_4=\frac{B_0}{r^2}+D_0
\end{equation*}
and we can integrate (2.2)-(2.4) to obtain
\begin{align}
G_1(r,\theta)&=-\frac{B_0\dot{S}}{r}+\beta(\theta)\\
G_2(r,\theta)&=2B_0\Big(R+\frac{S}{r^2}\Big)-\frac{B_0}{2r^2}\ddot{S}+\frac{\dot{\beta}}{r}+\xi(\theta)
\end{align}
where $\beta(\theta)$ was introduced in (2.14) and $\xi(\theta)$ appears as an integration constant (an arbitrary function of $\theta$). Equations (2.2) to (2.4) further imply
\begin{equation}
\ddot{\beta}+\beta=0,\quad \dot{\xi}-3D_0\dot{S}=0,\quad B_0(S^{(3)}-4\dot{S})=0
\end{equation}
and the integral (2.9) reduces to 
\begin{equation}
Y=D_0L_3^3+B_0L_3(p_1^2+p_2^2)+\{g_1(x,y),p_1\}+\{g_2(x,y),p_2\}
\end{equation}

(ii) From now on we assume that at least one of the constants $A_i,B_j,C_j$ does not vanish. From (\ref{RAi})-(\ref{RAA4C2}) we see that the most general form of the radial term $R(r)$ is 
\begin{align}
 R(r)&=\frac{a_1}{r^4}+\frac{a_2}{r^3}+\frac{a_3}{r}+a_4r^2+a_5r^4+a_6\log r\nonumber\\
 &+\frac{a_7}{\sqrt{A+Cr^2}}+\frac{a_8}{\sqrt{A+Cr^2}}\log\Bigg(\frac{\sqrt{A}+\sqrt{A+Cr^2}}{r}\Bigg)
 \end{align} 
 where $(A,C)=(A_3,C_1)$ or $(A_4,C_2)$.
 
 Let us show that the ``exotic'' tems $a_1,a_2,a_5,a_6,a_7$ and $a_8$ are actually absent, i.e. their presence is not allowed by the original determining equations (\ref{detunpol})-(\ref{detquatrepol}). We proceed systematically by assuming the contrary.\\
 
 (ii.1) $a_1\neq0$.

The function $R(r)$ must have the form (3.11) and the only possible nonzero constants  are  $C_1$ and $C_2$ (and as always $B_0$ and $D_0$).

From (\ref{fonctionun}) to (\ref{fonctionquatre}),  we find
\begin{eqnarray}
F_1=0,\quad F_2=B_0,\quad F_3=C_1\cos\theta+C_2\sin\theta,\nonumber\\ F_4=\frac{B_0}{r^2}+\frac{1}{r}(-C_1\sin\theta+C_2\cos\theta)+D_0
\end{eqnarray}

The $r^0$ term in the compatibility condition (2.10) is
 \begin{equation}
 a_1(C_1\cos\theta+C_2\sin\theta)=0
 \end{equation}

The condition $C_1^2+C_2^2\neq0$ implies $a_1=0$.\\

(ii.2) $a_1=0$, $a_2\neq0$.

The function $R(r)$ must have the form (\ref{RBi}) and the possible nonzero constants in $Y$ are $B_1,B_2,C_1,C_2$ (in addition to $B_0$ and $D_0$).

The coefficient of $r^0$ in (\ref{compa}) is
 \begin{equation}
 a_2(B_1\cos2\theta+B_2\sin2\theta)=0
 \end{equation}
We can keep $a_2\neq0$ only if we impose $B_1=B_2=0$.

The coefficient of $r^0$ in (\ref{detunpol}) is
\begin{equation}
a_2B_0\dot{S}=0
\end{equation}

 For $B_0=0$ or $\dot{S}=0$, the term in $r^1$ in (2.10) implies 
 \begin{equation}
 a_2(C_1\cos\theta+C_2\sin\theta)=0
 \end{equation}
a contradiction with the assumption $C_1^2+C_2^2\neq0$ (since we already have $B_1=B_2=0$). Thus we have $a_2=0$.\\

(ii.3) $a_1=a_2=0$, $a_5\neq0$.

The function $R(r)$ satisfies (\ref{RAi}) and we impose $A_1^2+A_2^2\neq0$. (\ref{G1}) and (\ref{detquatrepol}) give the explicit form of the functions $G_1$ and $G_2$ and from the term $r^{12}$ in  (\ref{detunpol}) we obtain
 \begin{equation}
 a_5^2(A_1\cos3\theta+A_2\sin3\theta)=0
 \end{equation}
a contradiction. Hence we have $a_5=0$.\\

(ii.4) $a_1=a_2=a_5=0$, $a_6\neq0$.

The function $R(r)$ must have the form (\ref{RAAi}) and we request $A_3^2+A_4^2\neq0$.  We obtain from (\ref{G1}) and (\ref{detquatrepol}) the functions $G_1$ and $G_2$ and (\ref{detunpol}) in this case contains an   $r^4\log r$ term with coefficient
 \begin{equation*}
 a_6^2(A_3\cos\theta+A_4\sin\theta)=0
 \end{equation*}
and hence   $a_6=0$.\\

(ii.5) $a_1=a_2=a_5=a_6=0$, $a_8\neq0$.

The function $R(r)$ must have the form (\ref{RAA3C1}) or (\ref{RAA4C2}) and we request $A_3C_1\neq0$ or $A_4C_2\neq0$. Inserting $G_1$ and $G_2$ obtained from (\ref{G1}) and (\ref{detquatrepol}) in (\ref{detunpol}) we have the term  $\displaystyle r^{12}\log^2\Bigg(\frac{\sqrt{A}+\sqrt{A+Cr^2}}{r}\Bigg)$ with coefficient
\begin{equation}
 a_8^2AC^4\cos\theta=0
 \end{equation}
Since we impose $AC\neq0$, we find $a_8=0$.\\

(ii.6) $a_1=a_2=a_5=a_6=a_8=0$, $a_7\neq0$.

The radial part is
\begin{equation}
R(r)=\frac{a_3}{r}+\frac{a_7}{\sqrt{A+Cr^2}}
\end{equation}
and we must request that $AC\neq0$. From the coefficients of  $r^9$ and $r^7$ in (\ref{detunpol}) we have that:
\begin{eqnarray}
a_7C^3(B_0\dot{S}-2a_3A\cos\theta)=0\\
a_7AC^2(8B_0\dot{S}-15a_3A\cos\theta)=0
\end{eqnarray}

Then for any values of $B_0$, we have 
\begin{equation*}
a_3A\cos\theta=0
\end{equation*}
Since  $A\neq0$ we have $a_3=0$. In (\ref{detunpol}) we have an $r^2$ term with coefficient 
\begin{equation}
a_7A^4(\sin\theta\dot{S}+2\cos\theta S)=0 
\end{equation}
and hence $\dot{S}=-2\cot\theta S$. Substituting $\dot{S}$ into (\ref{detunpol}) we use the coefficient of $r^{10}$ to obtain
\begin{equation}
\beta(\theta)=\frac{C\cos\theta}{4}(-\hbar^2+8S)
\end{equation}
Then  the coefficient of $r^8$ gives
\begin{equation}
a_7AC^2\cos\theta=0
\end{equation}
Since $AC\neq0$ it follows that $a_7=0$.

(iii) So far we have shown  that the function $R(r)$ must have the form 
\begin{equation}
R(r)=\frac{a_3}{r}+a_4r^2
 \end{equation}
To complete the proof of the theorem we must show that either $a_3$ or $a_4$ must vanish. From (\ref{RAi})-(\ref{RAA4C2}) we see that the constants in  $Y$ that can survive  (in addition to $D_0$ and $B_0$) are $A_1,A_2,A_3,A_4,B_1,B_2$. From (\ref{detunpol}) we obtain the coefficient of $r^8$ to be
\begin{equation}
a_4^2(A_1\cos3\theta+A_2\sin3\theta+A_3\cos\theta+A_4\sin\theta)=0
 \end{equation}
and hence for  $a_4\neq0$, we have  $A_1=A_2=A_3=A_4=0$.

The coefficient of $r^6$ yields
\begin{equation}
a_4\beta=0
\end{equation}
and hence $\beta=0$. Taking this into account we return to (\ref{compa}) and find the coefficient of  $r^2$:
\begin{equation}
a_3(B_1\sin2\theta-B_2\cos2\theta)=0
 \end{equation}
For $a_3\neq0$ this implies $B_1=B_2=0$ and we are back in the generic case where the integral is (3.19).

Thus either $a_3$ or $a_4$ in (3.34) must vanish and this completes the proof of the theorem.
\end{proof}

We see that a third order integral $Y$ of (\ref{H}) with at least one nonzero constant $A_i, B_j, C_j$ with $i=1,2,3,4$, $j=1,2$ can only exist if the radial part of the potential is a harmonic oscillator $R(r)=ar^2$, a Coulomb-Kepler potential $R(r)=a/r$ or $R=0$.
\section{Angular term $\frac{S(\theta)}{r^2}$ in the potential}
Let us return to the problem of solving the determining equations (\ref{detunpol}) to (\ref{detquatrepol}) and concentrate on the angular part, once the radial part is known. We shall consider each of the four cases of Theorem 1 separately.

\subsection{Radial equations satisfied for all $R(r)$}
According to Theorem 1, the constants in the third order integral $Y$ satisfy (3.15) and $Y$ itself is as in  (3.19). In this case the functions $G_1(r,\theta)$ and $G_2(r,\theta)$ are as in (3.16) and (3.17) where $\beta(\theta)$, $\xi(\theta)$ and $S(\theta)$ satisfy (3.18).

Two cases must be considered separately:\\

1. $B_0\neq0$

From (3.18) we obtain
\begin{align}
\label{conditionsWei}
\xi(\theta)&=3D_0S+\xi_0\nonumber\\
\beta(\theta)&=\beta_1\cos\theta+\beta_2\sin\theta \nonumber\\
S(\theta)&=s_1\cos2\theta+s_2\sin2\theta+s_0
\end{align}
where $\xi_0, \beta_1$, $\beta_2, s_0, s_1, s_2$ are constants (we can put $s_0=0$). 

Substituing (4.1) in  (\ref{detunpol}) we obtain $s_1=s_2=0$, and we have a purely radial potential. In the integral (3.19) we have $g_1=g_2=0$ and the result is trivial. Namely, since $L_3$ is an integral, $L_3^3$ and $L_3H$ are also integrals. In general this potential is not superintegrable but first order integrable.\\

    2) $B_{0}=0$, $D_0=1$ 
     
The third equation in (3.18) is satisfied trivially  so $S(\theta)$ in (\ref{conditionsWei}) is arbitrary. Putting $\beta(\theta)$ and $\xi(\theta)$ in (\ref{detunpol}) we obtain
\begin{align}
\label{a300} (\beta_1\cos\theta + \beta_2\sin\theta)r^3R' =
&r(\frac{\hbar^2}{4}S^{(3)}-3S\dot{S}-\xi_0\dot{S})\nonumber\\&+\Big((\beta_1\sin\theta
- \beta_2\cos\theta)\dot{S}+2(\beta_1\cos\theta +
\beta_2\sin\theta)S\Big)
\end{align}
We distinguish two subcases:\\

2.a) $\beta_1=\beta_2=0$ \\
We have
\begin{equation*}
\label{A300} \hbar^2S^{(3)}=12S\dot{S}+4\xi_0\dot{S}
\end{equation*}
This can be  integrated to:
\begin{equation}
\label{elliptic} \hbar^2\dot{S}^2=4S^3+4\xi_0S^2+bS+c
\end{equation}
where $b$ and $c$ are integration constants. We can set $\xi_0=0$ by a suitable change of variables $S(\theta)\mapsto T(\theta)-\frac{\xi_0}{3}$ and (\ref{elliptic}) is simplified to
\begin{equation*}
\label{weirs} \hbar^2\dot{T}^2=4T^3-t_2T-t_3
\end{equation*}
where $t_2$ and $t_3$ are constants.

The potential is expressed in terms of the Weierstrass elliptic  function:
\begin{equation}
\label{weirpot}
V(r,\theta) = R(r)+\frac{\hbar^2\wp(\theta,t_2,t_3)}{r^2}
\end{equation}
for an arbitrary radial part $R(r)$.

This potential has a third and a second order constant of motion of the form:
\begin{align}
\label{thirddep}
Y&=2L_3^3 + \{L_3,3\hbar^2\wp(\theta)\}\\
\label{seconddep}
X&=L_3^2 + 2\hbar^2\wp(\theta)
\end{align}

The second order constant of motion is known as the one-dimensional Lam\'e operator. $Y$ and $X$ are algebraically related \cite{Hiet2}:
\begin{equation}
\label{algrel}
\bigg(\frac{Y}{2}\bigg)^2=8\bigg(\frac{X}{2}\bigg)^3-\frac{1}{4}\hbar^4t_2X+\frac{1}{4}\hbar^6t_3
\end{equation}

This is a good example of what is called algebraic integrability \cite {Hiet1,Hiet2}. Since we are looking for hamiltonians with algebraically independent second and third order integrals of motion, this result is not a real superintegrable system. 

The classical potential corresponding to (\ref{weirpot})
is  a purely radial potential $V(r)=R(r)$. This result is directly obtained from (\ref{a300}) in the limit $\hbar\mapsto0$.\\

2.b) $\beta_1^2+\beta_2^2\neq0$ \\
Differentiating (\ref{a300}) two times with respect to  $r$, we obtain:
\begin{eqnarray*}
\Big(r^3R'\Big)''=0
\end{eqnarray*}
and hence
\begin{eqnarray*}
R(r)=\frac{a}{r}+\frac{b}{r^2}
\end{eqnarray*}
The $ \frac{1}{r^2}$ term can  be included in the angular part of the general form of the potential. So without loss of generality, we can set $b=0$. This case will be studied in section 4.2 since the radial part of the potential is of the form of case \emph{(ii)} of Theorem 1.

\subsection{Potential of the form $\displaystyle V(r,\theta)=\frac{a}{r}+\frac{S(\theta)}{r^2}$ with $a\neq0$}
We consider (\ref{detdeuxpol}) and (\ref{detquatrepol}) separate different powers of $r$ and obtain  equations relating $\beta(\theta)$ in (\ref{G1})  with the angular part $S(\theta)$ of the potential: 
\begin{align}
 &\ddot{\beta}+\beta -2(C_1\cos\theta+C_2\sin\theta)\ddot{S}+5(C_1\sin\theta-C_2\cos\theta)\dot{S}\nonumber\\
&+2(C_1\cos\theta+C_2\sin\theta)S=6a(B_1\sin2\theta-B_2\cos2\theta)\\\nonumber\\
&(B_1\cos2\theta+B_2\sin2\theta+B_0)S^{(3)}+8(-B_1\sin2\theta+B_2\cos2\theta)\ddot{S}\nonumber\\
&+4(-5B_1\cos2\theta-5B_2\sin2\theta+B_0)\dot{S}+16(B_1\sin2\theta-B_2\cos2\theta)S\nonumber\\
&=3a(-15A_1\cos3\theta-15A_2\sin3\theta+A_3\cos\theta+A_4\sin\theta)
\end{align}
\begin{align}
&(3A_1\sin3\theta-3A_2\cos3\theta+A_3\sin\theta-A_4\cos\theta)S^{(3)}\nonumber\\
&+(36A_1\cos3\theta+36A_2\sin3\theta+4A_3\cos\theta+4A_4\sin\theta)\ddot{S}\nonumber\\
&-(132A_1\sin3\theta-132A_2\cos3\theta-4A_3\sin\theta+4A_4\cos\theta)\dot{S}\nonumber\\
&-(144A_1\cos3\theta+144A_2\sin3\theta-16A_3\cos\theta-16A_4\sin\theta)S=0
\end{align}

Integrating (\ref{detquatrepol}) for $G_2$ we obtain the function \begin{equation*}\xi(\theta)=3D_0S(\theta)-a(C_1\sin\theta-C_2\cos\theta)+\xi_0\end{equation*}

From (\ref{detunpol}), it follows that $S(\theta)$ and $\beta(\theta)$  verify: 
\begin{align}
 &\hbar^2\big(D_0S^{(3)}+aC_1\cos\theta+aC_2\sin\theta\big)-4\xi\dot{S}+4a\beta=0\\\nonumber\\
&\hbar^2\bigg((-C_1\sin\theta+C_2\cos\theta)S^{(3)}-4(C_1\cos\theta+C_2\sin\theta)\ddot{S}\nonumber\\
&+6(C_1\sin\theta-C_2\cos\theta)\dot{S}+4(C_1\cos\theta+C_2\sin\theta)S\nonumber\\
&+6a(-B_1\sin2\theta+B_2\cos2\theta)\bigg)=4\dot{S}\dot{\beta}-8S\beta\nonumber\\
&-8(C_1\cos\theta+C_2\sin\theta)\dot{S}^2+12a(B_1\cos2\theta+B_2\sin2\theta+B_0)\dot{S}\nonumber\\
&-12a^2(A_1\cos3\theta+A_2\sin3\theta+A_3\cos\theta+A_4\sin\theta)\\\nonumber\\
&\hbar^2\bigg((B_1\cos2\theta+B_2\sin2\theta-B_0)S^{(3)}+8(-B_1\sin2\theta+B_2\cos2\theta)\ddot{S}\nonumber\\
&-4(5B_1\cos2\theta+5B_2\sin2\theta+B_0)\dot{S}+16 (B_1\sin2\theta-B_2\cos2\theta)S\bigg)\nonumber\\
&=-3a\hbar^2(5A_1\cos3\theta+5A_2\sin3\theta+A_3\cos\theta+A_4\sin\theta)\nonumber\\
&+12a(3A_1\sin3\theta-3A_2\cos3\theta+A_3\sin\theta-A_4\cos\theta)\dot{S}\nonumber\\
&+36a(A_1\cos3\theta+A_2\sin3\theta+A_3\cos\theta+A_4\sin\theta)S\nonumber\\
&+2(B_1\cos2\theta+B_2\sin2\theta+B_0)\dot{S}\ddot{S}-12(B_1\sin2\theta-B_2\cos2\theta)\dot{S}^2\nonumber\\
&-16(B_1\cos2\theta+B_2\sin2\theta+B_0)S\dot{S}
\end{align}

\begin{align}
&\hbar^2\bigg((3A_1\sin3\theta-3A_2\cos3\theta-3A_3\sin\theta+3A_4\cos\theta)S^{(3)}\nonumber\\
&+(36A_1\cos3\theta+36A_2\sin3\theta-12A_3\cos\theta-12A_4\sin\theta)\ddot{S}\nonumber\\
&(-132A_1\sin3\theta+132A_2\cos3\theta-12A_3\sin\theta+12A_4\cos\theta)\dot{S}\nonumber\\
&(-144A_1\cos3\theta-144A_2\sin3\theta-48A_3\cos\theta-48A_4\sin\theta)S\bigg)=\nonumber\\
&-72(A_1\cos3\theta+A_2\sin3\theta+A_3\cos\theta+A_4\sin\theta)S^2\nonumber\\
&+(-120A_1\sin3\theta+120A_2\cos3\theta-40A_3\sin\theta+40A_4\cos\theta)S\dot{S}\nonumber\\
&+(54A_1\cos3\theta+54A_2\sin3\theta+6A_3\cos\theta+6A_3\sin\theta)\dot{S}^2\nonumber\\
&+(6A_1\sin3\theta-6A_2\cos3\theta+2A_3\sin\theta-2A_4\cos\theta)\dot{S}\ddot{S}
\end{align}

The principal result that follows from the compatibility of the preceding determining equations is that the only potential that satisfies (4.8) to (4.14) is
\begin{equation}
\label{smoro2}
V (r,\theta) =\displaystyle\frac{a}{r}+\frac{\alpha_1+\alpha_2\sin\theta}{r^2\cos^2\theta}
\end{equation}
or potentials that can be rotated to (\ref{smoro2}).

The potential (\ref{smoro2}) is a well known quadratically superintegrable one \cite{winfris1,winfris2} with the Coulomb potential as a special case. The third order integral is the commutator  (or Poisson commutator) of the  two second order ones. For more recent discussions of the potential (\ref{smoro2}) see e.g. \cite{shef1,ttw1,grav,let}.

\subsection{Potential of the form $\displaystyle V(r,\theta)=ar^2+\frac{S(\theta)}{r^2}$ with $a\neq0$}
Here, from Theorem 1, the third order constant of motion is of the form:
\begin{align}
 Y &= D_{0}L_3^3 + B_{0}L_3(p_1^2+p_2^2)+B_1\{L_3,p_1^2-p_2^2\}+B_2\{L_3,p_1p_2\}\nonumber\\
&+\{g_1,p_1\}+\{g_2,p_2\}
\end{align}
As in the preceding case, we obtain from (\ref{detdeuxpol}) to (\ref{detquatrepol}) that the angular part of the potential has to be a solution of:
\begin{align}
 &(B_1\cos2\theta+B_2\sin2\theta+B_0)S^{(3)}+8(-B_1\sin2\theta+B_2\cos2\theta)\ddot{S}\nonumber\\
&+4(-5B_1\cos2\theta-5B_2\sin2\theta+B_0)\dot{S}+16 (B_1\sin2\theta-B_2\cos2\theta)S=0
\end{align}
for $\xi(\theta)=3D_0S+\xi_0$ and from (\ref{detunpol}):
\begin{align}
&\hbar^2D_0S^{(3)}-12D_0S\dot{S}-4\xi_0\dot{S}=0\\\nonumber\\
&\hbar^2\bigg((B_1\cos2\theta+B_2\sin2\theta-B_0)S^{(3)}+8(-B_1\sin2\theta+B_2\cos2\theta)\ddot{S} \nonumber\\
&-4(5B_1\cos2\theta+5B_2\sin2\theta+B_0)\dot{S}+16 (B_1\sin2\theta-B_2\cos2\theta)S\bigg) \nonumber\\
&=2(B_1\cos2\theta+B_2\sin2\theta+B_0)\dot{S}\ddot{S}-12(B_1\sin2\theta-B_2\cos2\theta)\dot{S}^2 \nonumber\\
&-16(B_1\cos2\theta+B_2\sin2\theta+B_0)S\dot{S}
\end{align}
for $\beta(\theta)=0$.

We can solve the compatibility between all those determining equations for $S$ to obtain the following  potential:
\begin{align}
\label{smoro1}
V (r,\theta) &=ar^2+\frac{2(b+c)+2(c-b)\cos2\theta}{r^2\sin^22\theta}\nonumber\\&=ar^2+\frac{b}{x^2}+\frac{c}{y^2}
\end{align}
or potentials that can be rotated to (\ref{smoro1}).

The potential (\ref{smoro1}) is also a well known quadratically superintegrable potential having the harmonic oscillator as a special case \cite{winfris1,winfris2}. The third order integral can be obtained as a commutator of the second order ones.
\subsection{Potential of the form $\displaystyle V(r,\theta)=\frac{S(\theta)}{r^2}$ }

 From (\ref{detunpol}) to (\ref{detquatrepol}), we again obtain the determining equations (4.8) to (4.14) with $a=0$. The third order constant of motion is in its most general form (\ref{intmod}).

Solving the determining equations for $S(\theta)$ and $\beta(\theta)$ (with $a=0$), we reobtain special cases of results derived in sections 4.2 and 4.3 (without the Coulomb or the harmonic radial parts). In addition to these known cases we obtain three additional ones. The first is
\begin{equation}
\label{calo}
V(r,\theta)=\frac{\alpha}{r^2\sin^23\theta}
\end{equation}
This potential is a special case of the rational three-body  Calogero system in two dimensions and  is already known to be superintegrable \cite{wo}.

 The third order constant of motion associated to (\ref{calo}) is:
 \begin{equation*}
 Y=p_1^3-3p_1p_2^2+2\alpha\bigg\{p_1,\frac{-3x^4+6x^2y^2+y^4}{y^2(-3x^2+y^2)^2}\bigg\}+\alpha\bigg\{p_2,\frac{16xy}{(-3x^2+y^2)^2}\bigg\}
 \end{equation*}
 
The potential is obtained both in classical and quantum mechanics. 

The two others cases occur  when the third order integral of motion takes the form
 \begin{equation}
\label{intc1d0} 
Y=C_1\{L_3^2,p_1\}+2D_0L_3^3+\{g_1,p_1\}+\{g_2,p_2\}
 \end{equation}
 i.e. $A_1=A_2=A_3=A_4=B_0=B_1=B_2=C_2=0$. In this case, (\ref{detunpol}) to (\ref{detquatrepol}) are reduced to the system for $S(\theta)$ and $\beta(\theta)$:
\begin{gather}
\label{S1beta}
\ddot{\beta}+\beta+ C_1(-2\cos\theta\ddot{S}+5\sin\theta\dot{S}+2\cos\theta S)=0\\
\hbar^2D_0S^{(3)}-4\xi\dot{S}=0\\
\label{S1C1}
 \hbar^2C_1\bigg(-\sin\theta S^{(3)}-4\cos\theta\ddot{S}+6\sin\theta\dot{S}
+4\cos\theta S\bigg)=4\dot{S}\dot{\beta}-8S\beta-8C_1\cos\theta\dot{S}^2
\end{gather}
with $\xi(\theta)=3D_0S+\xi_0$.

We distinguish two subcases:\\

4.4.a) $D_0=0$ and $C_1=1$. 

 In this case, we have $\xi=0$ or $S=0$. If we set  $\xi=0$ and $S=\dot{T}$, (\ref{S1beta}) is solved directly for $\beta$
\begin{equation}
\label{betaC1}
\beta(\theta)=\beta_1\cos\theta+\beta_2\sin\theta-\sin\theta+2\cos\theta\dot{T}
\end{equation}

Inserting (\ref{betaC1}) in (\ref{S1C1}) we obtain a fourth order ODE for T:
\begin{align}
\label{T4th}
&\hbar^2\big(\sin\theta T^{(4)}+4\cos\theta T^{(3)}-6\sin\theta\ddot{T}-4\cos\theta\dot{T}\big)-12\sin\theta\dot{T}\ddot{T}\nonumber\\
&-4\cos\theta T\ddot{T}-4\big(\beta_1\sin\theta-\beta_2\cos\theta\big)\ddot{T}-16\cos\theta\dot{T}^2 \nonumber\\
&+8\sin\theta T\dot{T}-8\big(\beta_1\cos\theta+\beta_2\sin\theta\big)\dot{T}=0
\end{align}

Under the transformation $(\theta,T(\theta))\mapsto(z,T(z))$ where $z=\tan\theta$ (\ref{T4th}) becomes:
\begin{align}
\label{T4thz}
& \hbar^2z(1+z^2)^2T''''+4\hbar^2(1+z^2)(1+3z^2)T'''+\big[2\hbar^2z(13+18z^2)-4\beta_1 z+4\beta_2 \nonumber\\
&-4T-12z(1+z^2)T')\big]T''
-8(2+3z^2)T'^2-8\beta_1 T'+4\hbar^2(1+6z^2)T'=0
\end{align}

This equation  can be integrated twice. The first integral is:
\begin{align}
\label{T3thz}
&\hbar^2z^2(1+z^2)^2T'''+2\hbar^2z(1+z^2)(1+3z^2)T''-6z^2(1+z^2)T'^2+2(-\hbar^2+\hbar^2z^2\nonumber\\
&+3\hbar^2z^4-2\beta_1z^2+2\beta_2z-2zT)T'+2T^2-4\beta_2T=K_1
\end{align}
for an arbitrary constant of integration $K_1$.

The second integral is:
\begin{align}
\label{T2thz}
&\bigg[T''+\frac{z(2z^2+1)T'-T+\beta_2}{z^2(z^2+1)}\bigg]^2=
\frac{1}{\hbar^2z^4(z^2+1)^3}\bigg[4z^4(z^2+1)^2(T')^3\nonumber\\
&+z^2(z^2+1)\big[4zT+\hbar^2(2z^2+1)+4\beta_1z^2-4\beta_2z\big](T')^2-2z(z^2+1)\big[2zT^2\nonumber\\
&-(4\beta_2z+\hbar^2)T-(K_1z-\beta_2\hbar^2)\big]T'-\big[4zT^3+\big(\hbar^2(z^2-1)+4\beta_1z^2-12\beta_2z\big)T^2\nonumber\\
&-2\big(K_1z+\beta_2(4\beta_1+\hbar^2)z^2-4\beta_2^2z-\beta_2\hbar^2\big)T-(\hbar^2K_2z^2-2\beta_2K_1z+\beta_2^2\hbar^2)\big]\bigg]
\end{align}
for a second arbitrary constant of integration $K_2$.

The transformation $(z,T(z))\mapsto(x,W(x))$:
\begin{equation}
 z=\frac{2\sqrt{x}\sqrt{1-x}}{1-2x},\quad T=\frac{8\hbar^2W+(\hbar^2+4\beta_1)(1-2x)}{8\sqrt{x}\sqrt{1-x}}+\beta_2
\end{equation}
maps (\ref{T4thz}), (\ref{T3thz}) and (\ref{T2thz})  to equations  contained in a series of papers by C. Cosgrove on higher order Painlev\'e equations. Specifically, (\ref{T4thz}) is mapped into the fourth order equation F-VII \cite{cos1} (see also \cite{cos2}). Equation (\ref{T3thz}) is mapped into the third order differential equation Chazy-I.a  of \cite{cos3} and (\ref{T2thz}) into the second order differential equation of second degree  SD-I.a of \cite{cos4} with parameters
\begin{gather}
c_1=c_4=c_5=c_6=c_8=0,\quad c_2=-c_3=1,\quad c_7=\frac{\hbar^2+8\beta_1}{16\hbar^2} \\
c_9=\frac{\hbar^4-16\beta_1^2-16\beta_2^2-8K_1}{64\hbar^4} \\
c_{10}=-\frac{16\hbar^2K_2-8(4\beta_1+\hbar^2)K_1+\hbar^2(4\beta_1+\hbar^2)^2}{256\hbar^6} 
\end{gather}

SD-I.a is the first canonical subcase of the more general equation that Cosgrove called the "master Painlev\'e equation``, SD-I \cite{cos4}. SD-I.a is solved by the Backlund correspondence
\begin{align}
 W(x)&=\frac{x^2(x-1)^2}{4P_6(P_6-1)(P_6-x)}\bigg[P_6'-\frac{P_6(P_6-1)}{x(x-1)}\bigg]^2+\frac{1}{8}(1-\sqrt{2\gamma_1})^2(1-2P_6)\nonumber\\
&-\frac{1}{4}\gamma_2\bigg(1-\frac{2x}{P_6}\bigg)-\frac{1}{4}\gamma_3\bigg(1-\frac{2(x-1)}{P_6-1}\bigg)+\bigg(\frac{1}{8}-\frac{\gamma_4}{4}\bigg)\bigg(1-\frac{2x(P_6-1)}{P_6-x}\bigg)
\end{align}
and 
\begin{align}
 W'(x)=-\frac{x(x-1)}{4P_6(P_6-1)}\bigg[P_6'-\sqrt{2\gamma_1}\frac{P_6(P_6-1)}{x(x-1)}\bigg]^2-\frac{\gamma_2(P_6-x)}{2(x-1)P_6}-\frac{\gamma_3(P_6-x)}{2x(P_6-1)}
\end{align}
where $\sqrt{2\gamma_1}$ can take either sign and  $\gamma_1,\gamma_2,\gamma_3$ and $\gamma_4$ are the arbitrary parameters that define the sixth Painlev\'e transcendent $P_6$ obtained from the well known second order differential equation:
\begin{eqnarray}
\label{P6}
 P_6''=\frac{1}{2}\bigg[\frac{1}{P_6}+\frac{1}{P_6-1}+\frac{1}{P_6-x}\bigg](P_6')^2-\bigg[\frac{1}{x}+\frac{1}{x-1}+\frac{1}{P_6-x}\bigg]P_6'\nonumber\\
+\frac{P_6(P_6-1)(P_6-x)}{x^2(x-1)^2}\bigg[\gamma_1+\frac{\gamma_2x}{P_6^2}+\frac{\gamma_3(x-1)}{(P_6-1)^2}+\frac{\gamma_4x(x-1)}{(P_6-x)^2}\bigg]
\end{eqnarray}

The parameters $\gamma_1,\gamma_2,\gamma_3$ and $\gamma_4$ are related to the arbitrary constants of integration $\beta_1,\beta_2,K_1$ and $K_2$ 
\begin{align}
&-4c_7= \gamma_1-\gamma_2+\gamma_3-\gamma_4-\sqrt{2\gamma_1}+1,\\
\label{rela}
&-4c_{8}=0=(\gamma_2+\gamma_3)(\gamma_1+\gamma_4-\sqrt{2\gamma_1}),\\
&-4c_9=(\gamma_3-\gamma_2)(\gamma_1-\gamma_4-\sqrt{2\gamma_1}+1)+\frac{1}{4}(\gamma_1-\gamma_2-\gamma_3+\gamma_4-\sqrt{2\gamma_1})^2,\\
&-4c_{10}=\frac{1}{4}(\gamma_3-\gamma_2)(\gamma_1+\gamma_4-\sqrt{2\gamma_1})^2+\frac{1}{4}(\gamma_2+\gamma_3)^2(\gamma_1-\gamma_4-\sqrt{2\gamma_1}+1)  
\end{align}
 
Only three parameters of (\ref{P6}) are arbitrary in our case. From (\ref{rela}), we see that one of the following relations must hold
\begin{equation}
\label{relations}
 \gamma_2=-\gamma_3,\quad \gamma_4=-\gamma_1+\sqrt{2\gamma_1}.
\end{equation}
From the inverse transformation $x\to z=\tan\theta$ 
\begin{equation}
x_{\pm}=\frac{1}{2}\pm\frac{1}{2\sqrt{1+z^2}}  =\Bigg\{\begin{array}{ll}
    \sin^2\big(\frac{\theta}{2}\big)\\\\
    \cos^2\big(\frac{\theta}{2}\big)\\
    \end{array}
\end{equation}
we obtain two solutions  for $S$. By taking the derivative of $T$ in (4.31) we obtain the quantum potentials 
\begin{eqnarray}
 V(r,\theta)=\frac{1}{r^2}\Bigg(\hbar^2W'(x_{\pm})-\frac{\pm8\hbar^2\cos\theta W(x_{\pm})+4\beta_1+\hbar^2}{4\sin^2\theta}\Bigg)
\end{eqnarray}

In the limit $\hbar\mapsto0$, (\ref{T3thz}) is reduced to a first order differential equation of second degree in $T'$:
\begin{eqnarray}
\label{class1}
3z^2(1+z^2)T'^2+2zTT'-T^2+2(\beta_1z^2-\beta_2z)T'+2\beta_2T+\frac{K_1}{2}=0
\end{eqnarray}

(\ref{class1}) is a special case of the more general equation:
\begin{equation}
\label{mitri}
 A(z)T'^2+2B(z)TT'+C(z)T^2+2D(z)T'+2E(z)T+F(z)=0
\end{equation}

A number of papers has been devoted to the integration of (\ref{mitri}). For example in \cite{mitri} a method is suggested for its integration.

Special solutions can be obtained under the condition that 
\begin{equation}
\label{condition}
\left| \begin{array}{ccc}
 A & B & D \\
B & C & E\\
D & E & F \end{array} \right|
=0
\end{equation}

This condition implies that $\beta_1=0$ and $K_1=-2\beta_2^2$ in (\ref{class1}). In this case (\ref{class1}) can be factorized 
\begin{equation}
\Big(\big(z+z\sqrt{4+3z^2}\big)T'-T+\beta_2\Big)\Big(\big(z-z\sqrt{4+3z^2}\big)T'-T+\beta_2\Big)=0
\end{equation}

We obtain two solutions:
\begin{align}
T_1&=\beta_2+\alpha\frac{z^{\frac{1}{3}}(5+3z^2+2\sqrt{4+3z^2})^{\frac{1}{6}}}{(2+\sqrt{4+3z^2})^{\frac{2}{3}}}\\
T_2&=\beta_2+\alpha\frac{(1+z^2)^{\frac{1}{3}}(2+\sqrt{4+3z^2})^{\frac{2}{3}}}{z(5+3z^2+2\sqrt{4+3z^2})^{\frac{1}{6}}}
\end{align}
where $\alpha$ is an integration constant.
The angular part of the potential is obtained by differentiating the preceding results with respect to $\theta$ and the resulting classical potentials are
\begin{eqnarray}
 V=\frac{3\alpha\sec^4\theta\Big[7+3\sqrt{4+3\tan^2\theta}+\cos2\theta\big(1+\sqrt{4+3\tan^2\theta}\big)\Big]}{r^2\tan^{\frac{2}{3}}\theta\sqrt{4+3\tan^2\theta}\Big(2+\sqrt{4+3\tan^2\theta}\Big)^{\frac{5}{3}}\Big(5+3\tan^2\theta+2\sqrt{4+3\tan^2\theta}\Big)^{\frac{5}{6}}}\nonumber\\
\end{eqnarray}
and
\begin{eqnarray}
 V=-\frac{\alpha\sec^{\frac{2}{3}}\theta\Big[47+17\sqrt{4+3\tan^2\theta}+18\cot^2\theta(2+\sqrt{4+3\tan^2\theta})+3\tan^2\theta(5+\sqrt{4+3\tan^2\theta})\Big]}{2r^2\sqrt{4+3\tan^2\theta}\Big(2+\sqrt{4+3\tan^2\theta}\Big)^{\frac{1}{3}}\Big(5+3\tan^2\theta+2\sqrt{4+3\tan^2\theta}\Big)^{\frac{7}{6}}}\nonumber\\
\end{eqnarray}

If condition (\ref{condition}) is not satisfied, the general solution of (\ref{mitri}) is related to the general solution of the equation
\begin{equation}
\frac{dw}{dz}=\frac{M(z)w^3+N(z)w^2+P(z)w}{w^2+Q(z)}
\end{equation}
where
\begin{equation}
T(z)=\frac{w^2+mw+n}{pw}
\end{equation}
and $M,N,P$ and $Q$ are complicated algebraic expressions depending on $z$ and
\begin{eqnarray*}
 m=\frac{2(BD-AE)}{A\sqrt{B^2-AC}},\quad
n=\frac{-1}{A(B^2-AC)}\left| \begin{array}{ccc}
 A & B & D \\
B & C & E\\
D & E & F \end{array} \right|,\quad
p=\frac{2\sqrt{B^2-AC}}{A}
\end{eqnarray*}

4.4b) $D_0=1$ ($D_0\neq0$).

Equation (4.24) can be integrated to (4.3) so the angular part of the potential is expressed in terms of the Weierstrass elliptic function  $S(\theta)=\hbar^2\wp(\theta)$. The corresponding potential is
\begin{equation}
\label{weier2}
V(r,\theta)=\frac{\hbar^2\wp(\theta)}{r^2}
\end{equation}

From the compatibility of (4.23) and (4.25) we obtain 
\begin{align}
\label{betac1}
\beta(\theta)&=\frac{-C_1}{4 (h^2+\xi_0) \wp^2+4 b \wp+6 c}\Big[(b^2+5c(\hbar^2-2\xi_0))\cos\theta-2(3b-\hbar^4+\hbar^2\xi_0\nonumber\\
&+2\xi_0^2)\cos\theta\wp^2
-8(\hbar^2+\xi_0)\cos\theta\wp^3+b(\hbar^2+\xi_0)\sin\theta\dot{\wp}+(-2(9c+b(-\hbar^2\nonumber\\
&+\xi_0))\cos\theta+(3b+2\hbar^2(\hbar^2+\xi_0))\sin\theta\dot{\wp})\wp\Big]
\end{align}

The potential (\ref{weier2}) thus allows two third order integrals of motion (see (\ref{intc1d0}))
\begin{align}
\label{y1d0}
Y_1&=2L_3^3 + \{L_3,3\hbar^2\wp(\theta)\},\\
\label{y2c1} 
Y_2&=\{L_3^2,p_1\}+\{\beta\cos\theta+(2\cos\theta\dot{\wp}-\dot{\beta})\sin\theta,p_1\}\nonumber\\
&+\{\beta\sin\theta-(2\cos\theta\dot{\wp}-\dot{\beta})\cos\theta,p_2\}
\end{align}
where $\beta$ is the expression in (\ref{betac1}).

The integral (\ref{y1d0}) coincides with (\ref{thirddep}) for the more general potential (\ref{weirpot}). As noted  above (\ref{weirpot}) is not really superintegrable because of relation (\ref{algrel}). The potential (\ref{weier2}) is superintegrable since $Y_2$ (\ref{y2c1}), the second order integral of motion $X$ (\ref{seconddep}) and the Hamiltonian $H$ are algebraically independent.

In the classical limit $\hbar\mapsto 0$, the system  reduces to free motion.

\section{Summary and Conclusion}

The main results of this study can be summed up in two theorems.
\begin{theorem}
In classical mechanics in the Euclidean plane precisely 4 classes of Hamiltonian systems separating in polar coordinates and allowing a third order integral of motion exist. The corresponding potentials are (4.15), (4.20), (\ref{calo}) and 
\begin{equation}
V(r,\theta)=\frac{\dot{T}(\theta)}{r^2}
\end{equation}
where $T(z)$ satisfies equation (\ref{class1}) for $z=\tan\theta$. The third order integral of motion is 
\begin{equation}
 Y=\{L_3^2,p_1\}+\{\beta\cos\theta+(2\cos\theta\ddot{T}-\dot{\beta})\sin\theta,p_1\}\\
+\{\beta\sin\theta-(2\cos\theta\ddot{T}-\dot{\beta})\cos\theta,p_2\}
\end{equation}
\end{theorem}

The potential (4.15) and (4.20) are quadratically superintegrable and well-known. The third order integral is functionally dependent on the quadratic ones. The potential (\ref{calo}) is the three body Calogero system with no central term. Thus the genuinely new superintegrable classical potential is (5.1). We have not obtained the general solution of equation (\ref{class1}), but particular solutions led to the superintegrable potentials (4.51) and (4.52).

\begin{theorem}

In quantum mechanics the superintegrable systems correspond to the three known potentials (4.15), (4.20) and (\ref{calo}) plus two new ones. One new one is given by (4.44) where $W$ and $W'$ are expressed in terms of the sixth Painlev\'e transcendent $P_6$ in (4.35) and (4.36). The other new one is given by the Weierstrass elliptic functions $\wp(\theta)$ in  (4.55).
\end{theorem}

The Painlev\'e transcendents were first introduced in a study of movable singularities of second order nonlinear ordinary differential equations. They play an important role in the theory of classical infinite dimensional integrable systems.

The transcendent $P_6$ that was obtained as a superintegrable quantum potential in this article  depends on 3 free parameters (see (\ref{relations})). The Painlev\'e trancendents $P_1$, $P_2$ and $P_4$ have already appeared for potentials separable in cartesian coordinates \cite{gravel2}. A remarkable relation between quantum superintegrability and supersymmetry in quantum mechanics was discovered and used to solve the Schr\"odinger equation with potentials expressed in term of Painlev\'e transcendents \cite{mar3}-\cite{mar7}.

\section*{Acknowledgments}
We thank Professor C. M. Cosgrove for some valuable correspondence in which he generously helped us to solve equation (4.28). The research of P. W. was partially supported by NSERC of Canada.



\end{document}